\documentclass[preprint,12pt]{elsarticle}

\usepackage[margin=3cm]{geometry}
\usepackage{amssymb}
\usepackage{amsmath,amsfonts}
\usepackage{algorithmic}
\usepackage{algorithm}
\usepackage{array}
\usepackage[caption=false,font=normalsize,labelfont=sf,textfont=sf]{subfig}
\usepackage{textcomp}
\usepackage{stfloats}
\usepackage{url}
\usepackage{verbatim}
\usepackage{tabularray}

\def\figwidth{8.5cm}

\journal{Electirc Power System Research}

\begin{document}

\begin{frontmatter}

\title{Traveling Wave Method for Event Localization and Characterization of Power Transmission Lines}

\author[JSI,CS]{Marko Hudomalj\corref{cor}}
\ead{marko.hudomalj@ijs.si}
\author[FE]{Andrej Trost}
\author[JSI,CS]{Andrej Čampa}

\affiliation[JSI]{organization={Jožef Stefan Institute},
	city={Ljubljana},
	country={Slovenia}}

\affiliation[CS]{organization={Comsensus d.o.o},
	city={Dob},
	country={Slovenia}}

\affiliation[FE]{organization={University of Ljubljana, Faculty of Electrical Engineering},
	city={Ljubljana},
	country={Slovenia}}

\cortext[cor]{Corresponding author}

\begin{abstract}
Traveling wave theory is deployed today to improve the monitoring of transmission lines in electrical power grids. Most traveling wave methods require prior knowledge of the wave propagation of the transmission line, which is a major source of error as the value changes during the operation of the line. To improve the localization of events on transmission lines, we propose a new online localization method that simultaneously determines the frequency-dependent wave propagation characteristic from the traveling wave measurements of the event. Compared to conventional methods, this is achieved with one additional traveling wave measurement, but the method can still be applied in different measurement setups. We have derived the method based on the complex continuous wavelet transform. The accuracy of the method is evaluated in a simulation with a frequency-dependent transmission line model of the IEEE 39-bus system. The method was developed independently of the type of event and evaluated in test setups considering different lengths of the monitored line, line types and event locations. The localization accuracy is compared with existing online methods and analyzed with regard to the characterization capabilities. The method has a high relative localization accuracy in the range of 0.1\,\% under different test conditions.
\end{abstract}


\end{frontmatter}

\section{Introduction}
\label{sec:Introduction}
With the development of smart grids and the increasing complexity of electrical power systems, the need for advanced monitoring and decision-making is becoming increasingly important. This is especially true for transmission lines, which are crucial for grid operation but are typically long and often inaccessible and therefore difficult to monitor. In addition to established monitoring systems for slowly changing electrical properties, many efforts have recently been made to develop new systems that detect and respond to fast transient events \cite{schweitzerMillisecondMicrosecondNanosecond2016}. The methods for monitoring transient events are based on traveling wave (TW) theory \cite{wilches-bernalSurveyTravelingWave2021}. TW solutions on transmission lines are applied to different types of transient events: fault protection and localization \cite{aftabDynamicProtectionPower2020}, lightning strike detection \cite{liuHybridTravelingWave2009} and partial discharge (PD) localization and monitoring \cite{shafiqPerformanceComparisonPD2020, raoNewCrossCorrelationAlgorithm2020}.

\begin{figure}[H]
	\centering
	\includegraphics[width=\figwidth]{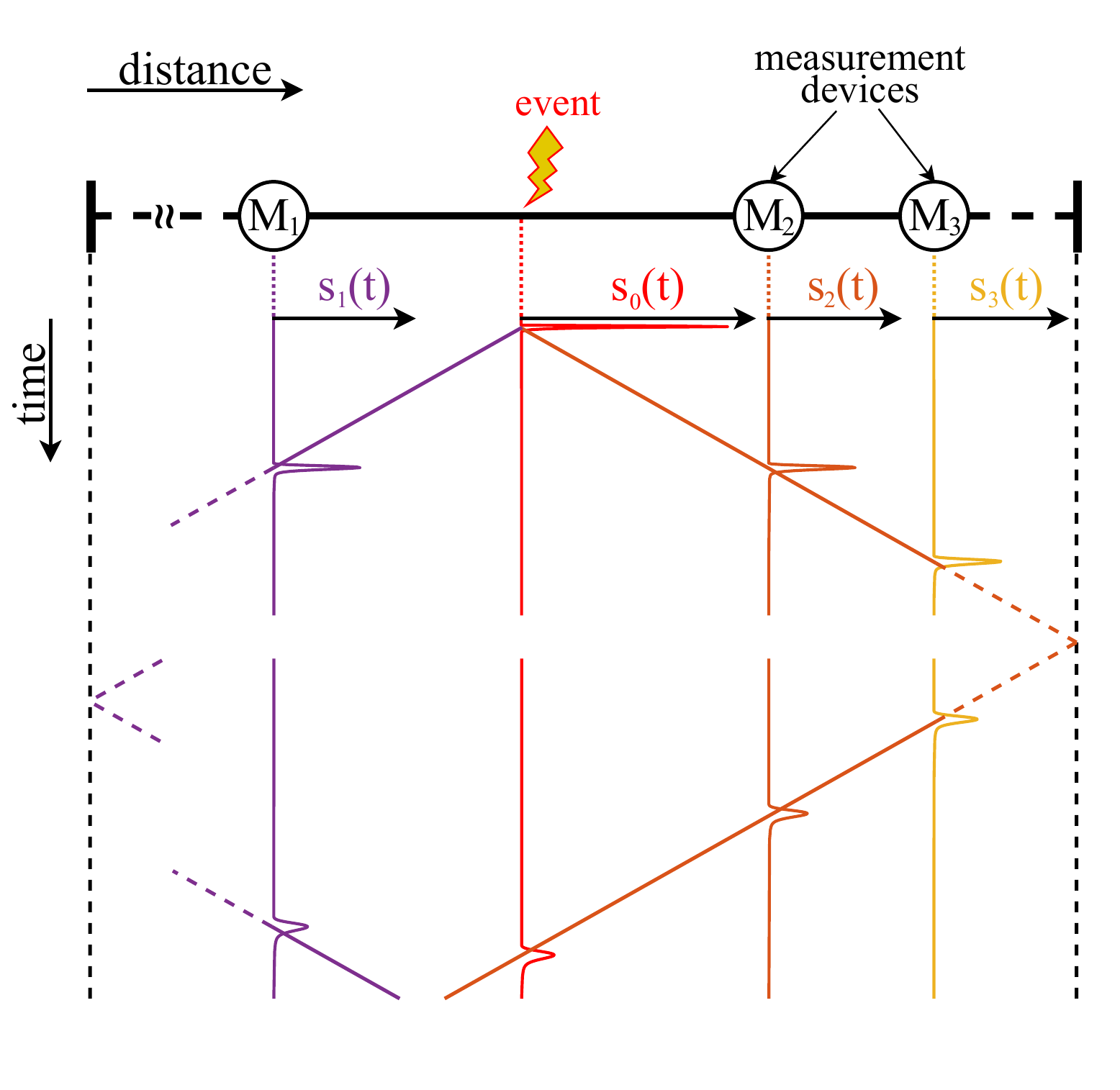}
	\caption{Lattice diagram of TW propagation on a transmission line with measurement devices. TW signals $s_i(t)$ are shown over the lattice diagram at the event location and the locations of measurement devices.}
	\label{fig:introduction_schematic}
\end{figure}

Transient events trigger TWs on transmission lines that propagate away from the source. Due to the frequency-dependent electromagnetic properties of the medium, the waves are attenuated and dispersed. The attenuation reduces the wave amplitude and dispersion elongates the wave in time, as propagation is frequency-dependent. Combined, attenuation and dispersion define the frequency-dependent wave propagation characteristic of a transmission line $\gamma(f)$. At the ends of the line where the medium changes, part of the wave is reflected and part is transmitted. Fig.~\ref{fig:introduction_schematic} shows a schematic overview of TW propagation on a transmission line. The TW is visualized with the lattice diagram, over which detailed waveforms are shown at the source and measurement locations. The TW propagates in both directions and is reflected at the line ends.

TW methods on transmission lines can be differentiated according to the number of measurement devices required, the need for time stamps and the consideration of wave propagation characteristic. Most methods have been developed for specific types of events, but in this paper we treat the methods independently of the event type.

Based on the number of devices, there are three main types \cite{aftabDynamicProtectionPower2020}: single-terminal \cite{raoNewCrossCorrelationAlgorithm2020, martinarroyoHalfSineMethodNew2019}, double-terminal \cite{mohamedPartialDischargeLocation2013, jiaImprovedTravelingWaveBasedFault2017} and multi-terminal \cite{chenComputationalFaultTime2019, mahdipourPartialDischargeLocalization2019}. The single-terminal methods require one measurement device at one end of the line, such as $M_1$ in Fig.~\ref{fig:introduction_schematic}. These methods are based on the detection of incident and reflected waves. The double-terminal type requires two measurement devices, one at each end of the monitored line, in Fig.~\ref{fig:introduction_schematic} represented by $M_1$ and $M_3$. In this case, both devices detect the incident waves. Multi-terminal methods work with several devices that can be positioned at the ends of the line branches \cite{chenComputationalFaultTime2019, robsonFaultLocationBranched2014} or on a single line \cite{mahdipourPartialDischargeLocalization2019}, as shown in Fig.~\ref{fig:introduction_schematic}. The advantage of a double-terminal setup is that it improves accuracy, but the disadvantage is the high requirements for time synchronization of the device \cite{liangPartialDischargeLocation2021}. Today, GPS synchronization meets most of the requirements \cite{mohamedPartialDischargeLocation2013}. On the other hand, the detection of reflected waves with single-terminal methods requires a high measurement resolution, as the waves are attenuated by the reflection and the waves travel longer on the line compared to the incident waves \cite{ragusaEffectRingMains2022}. It must be also assumed that the reflection does not change the wave phase and there are problems on branched networks, where reflected waves from different branches can precede each other and therefore cannot be distinguished. Some disadvantages can be improved with multi-terminal methods, but it must be assumed that the medium does not change in the different line sections. For this case, multi-terminal methods have been developed for hybrid transmission lines \cite{dengTravellingwavebasedFaultLocation2020}.

Most event localization techniques are based on time measurements by timestamping the TW. Traditional methods require two timestamps for event localization \cite{linTravellingWaveTimefrequency2012}. With these methods, the monitored line length and the $\gamma(f)$ in the form of the propagation velocity must be known in advance. Propagation velocity is also frequency dependent property $v(f)$ but in traditional methods the frequency dependence is neglected. Since noise and time synchronization can lead to errors in timestamping, a method based on cross-correlation of TW was proposed \cite{raoNewCrossCorrelationAlgorithm2020}, which, however, also assumes a known line length and constant velocity that neglects the dispersion effects.

In methods where $\gamma(f)$ or velocity must be known in advance, parameter can be determined based on transmission line modeling, using: geometry, conductor properties and soil resistance \cite{jiaImprovedTravelingWaveBasedFault2017}. However, it has been reported that more accurate results can be obtained by measurements \cite{shafiqIdentificationLocationPD2019}. The measurements can be carried out during commissioning or maintenance of the line, which usually requires the line to be de-energized. Even with measured parameters, significant errors can occur as the medium changes over time due to environmental changes and the operating condition of the line \cite{woutersThermalDistortionSignal2020}. In the case of overhead lines, the $\gamma(f)$ changes due to fluctuations in soil moisture and temperature \cite{liInfluenceFrequencyCharacteristics2016} and in the case of underground cables, the temperature changes caused by dynamic loading lead to fluctuations of $\gamma(f)$ in the range of 2\,\% \cite{vandeursenInfluenceTemperatureWave2021}.

Solutions for online monitoring have been proposed that do not require prior knowledge of $\gamma(f)$ or velocity. Some methods propose a signal generator that periodically generates a reference TW, on which the additional measurement is performed and $\gamma(f)$ is updated \cite{mahdipourPartialDischargeLocalization2019}. Similarly, a generator was deployed to determine the response of the line before and after a fault \cite{martinarroyoHalfSineMethodNew2019}. The fault is localized based on the difference in the online response. The disadvantage of using a reference signal generator is that additional devices must be installed and it is not suitable for all applications.

Most recent works have proposed online methods that do not require prior knowledge of $\gamma(f)$ and do not need a reference signal generator but additional measurements are need compared to conventional methods. One of these proposed methods uses two single-terminal devices at both ends of the line \cite{liangPartialDischargeLocation2021}. The devices operate as single-terminal event localization devices with initial assumption of propagation velocity. After both devices have localized an event, it is refined in iteration steps until the results from both ends match and the velocity is adjusted. However, the proposed method does not take into account the overall $\gamma(f)$, which is frequency-dependent, and uses timestamping of the TW in time where the peak of the wave is dependent on the dispersion. The method also requires the detection of multiple reflected waves, which has drawbacks as mentioned above.

Another source of error is the inaccurate setting of the line length as an input parameter for the traditional methods. The solution is to work in units per line length, which is then compared to a table of known feature locations on the transmission line \cite{kasztennyImprovingLineCrew2019a}. The location of the features can be determined during commissioning or maintenance by measurements or from past events where the location has been determined with great accuracy. The sections between the locations are then interpolated. The method uses two TW timestamps and compares them with a known TW line propagation time. The TW line propagation time is determined in advance based on measurements and indicates how long TW propagates from one end of the transmission line to the other, taking into account the wave propagation characteristic.

Table~\ref{tab:measurement_setups} summarizes the different measurement setups for event localization. The measurement setups are grouped according to the number of devices required and the way in which they consider $\gamma(f)$ or velocity. If the characteristic is entered as a parameter, only two TW measurements (incident and/or reflected waves) are required, but the accuracy of the method is affected by the variations in the medium. To account for the variations, three TW measurements are required, in this case no input parameters are needed. The table provides other requirements, assumptions and where the setups are most effective.

\begin{table}[H]
	\centering
	\small
	\caption{Comparison of event localization measurement setups.}
	\label{tab:measurement_setups}
	\SetTblrInner{rowsep=1pt}
	\begin{tblr}{c c c c c c c}
		\hline
		\SetCell[r=2]{m,1.2cm}Devices & \SetCell[c=4]{m} Requirements & & & & \SetCell[r=2]{m,2.5cm} Additional assumptions & \SetCell[r=2]{m,2cm} Effectiveness \\
		\hline
		&\SetCell[r=1]{m,2.2cm} Characteristic as parameter & Reflections & \SetCell[c=1]{m,1cm} Time Sync & \SetCell[c=1]{m,2cm} Grid Topology & & \\
		\hline
		\hline
		\SetCell[r=2]{m} 1 & / & 2 & No & Yes & \SetCell[r=3]{m,2.5cm} Reflection is measurable and not phase shifted & \SetCell[r=3]{m,2cm} Branchless topology \\
		\hline
		& $v(f)$ or $\gamma(f)$ & 1 & No & Yes & & \\
		\hline
		\SetCell[r=2]{m} 2 & / & 1 & Yes & Yes & & \\
		\hline
		& $v(f)$ or $\gamma(f)$ & 0 & Yes & No & & \SetCell[r=2]{m,1.5cm} Complex topology \\
		\hline
		3 & / & 0 & Yes & No & \SetCell[r=1]{m,2.5cm} Consistent transmission line in 2 sections & \\
		\hline
	\end{tblr}
\end{table}

To account for the changes in $\gamma(f)$ and thus improve event localization, we propose a new online method that determines the frequency-dependent wave propagation characteristic from TW measurements. Compared to traditional methods that require two TW measurements and a setting for $\gamma(f)$ or velocity, the proposed method requires three TW measurements and no setting, eliminating a major source of error. The proposed method provides results per line length and is not dependent on its accuracy. The method can be deployed in a single- or double-terminal measurement setup requiring reflected waves, or with three measurement devices in a multi-terminal setup requiring only incident waves, as presented in Table~\ref{tab:measurement_setups}. In this paper, we focus on the three-terminal solution as we consider it to be the most practical due to the problems with reflected waves already mentioned.

The theoretical basis of the method is derived in Section~\ref{sec:Theory} which is evaluated in a simulation setup described in Section~\ref{sec:Simulation}. The implementation of the method is described in Section~\ref{sec:Method} and the results are presented and discussed in Section~\ref{sec:Results}, including a comparison with other localization methods from the literature. Finally, conclusions are drawn in Section~\ref{sec:Conclusion}.

\section{Theory}
\label{sec:Theory}
\begin{figure}[H]
	\centering
	\includegraphics[width=\figwidth]{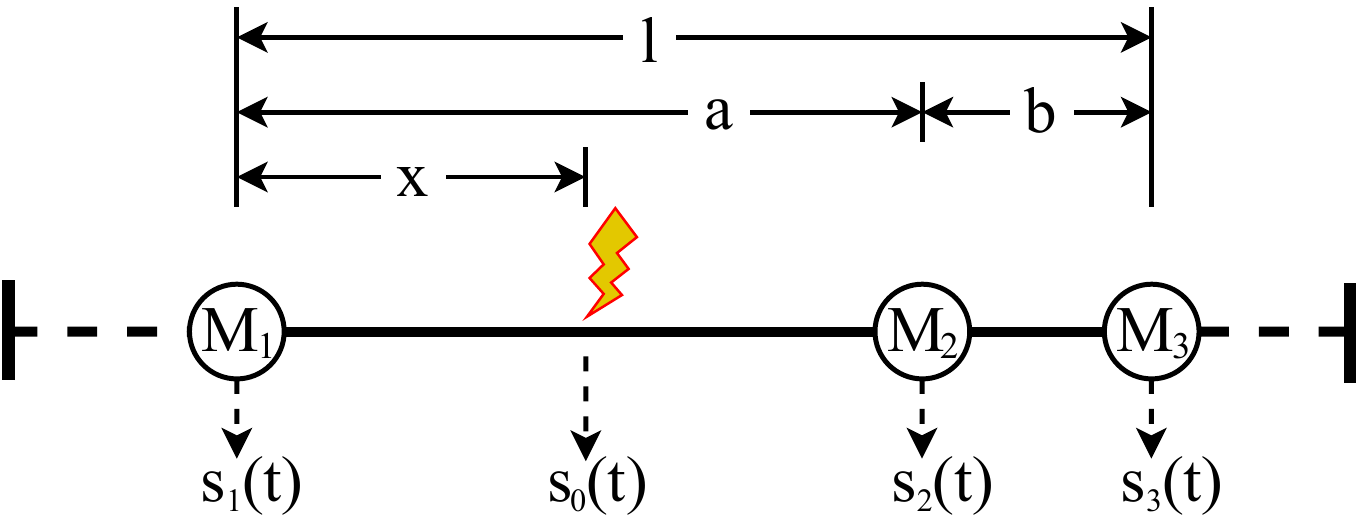}
	\caption{Diagram of a transmission line in 3-measurement devices setup with marked distances and considered time-series signals.}
	\label{fig:TL_diagram}
\end{figure}

In this section, we derive the analytical formulation of our method for event localization and wave propagation characterization. Our proposed approach is based on detected TWs that are the result of an event on the line. In this paper, we present the method based on measurements of three incident TWs, but the method can also be derived considering reflected waves. With reflected waves fewer measurement devices are needed but that comes with the disadvantages discussed in Table~\ref{tab:measurement_setups}. The measurement device setup with the signals and the distances considered for the analytical formulation is depicted in Fig.~\ref{fig:TL_diagram}.

The event location on the line is marked with the lightning bolt. The diagram shows the observed line segment with three measurement devices $M_1$, $M_2$ and $M_3$. The observed line is of length $l$ and the three measurement devices divide it in two sections. The event is located in one section of the observed line and the other section is event free. We set the reference location position at the location of measurement device $M_1$ from where the event is located at distance $x$ and the middle measurement device $M_2$ at distance $a$. The section between the measurement devices $M_2$ and $M_3$ is of length $b$. The length of the cable and the length between the measurement devices are not known exactly so the method uses lengths relative to the observed line length $l$ which are constant.

The line can extend beyond the devices as indicated by the dashed lines. Depending on how far the line extends beyond these two measurement devices, the time of the reflected TW and its magnitude changes. The signals at the measurement locations are denoted by $s_1$ to $s_3$, while $s_0$ is the signal at the event location.

Let us first consider the well-known forward TW propagation on a transmission line, as shown in Fig.~\ref{fig:introduction_schematic}, the waves propagate away from the event location to the measurement devices $M_1$, $M_2$ and $M_3$. In the frequency domain, the relation of signals at the source location and the measurement locations is:
\begin{equation}
	\begin{gathered}
		S_1(f) = e^{-x \gamma(f)} S_0(f) \\
		S_2(f) = e^{-(a-x) \gamma(f)} S_0(f) \\
		S_3(f) = e^{-(l-x) \gamma(f)} S_0(f)
	\end{gathered}
	\label{eqn:s1_s2_s3}
\end{equation}
where $S_0$ is the Fourier transform of the event signal and $S_1$,  $S_2$ and  $S_3$ are the Fourier transforms of the signals at the location of measurement devices at distances $x$, $a-x$ and $l-x$ away from the event origin. Transmission line is parametrised by complex and frequency-dependent wave propagation parameter $\gamma$:
\begin{equation}
	\gamma(f) = \alpha(f) + i\beta(f)
	\label{eqn:gamma_fun}
\end{equation}
where $\alpha$ and $\beta$ are frequency-dependent real and imaginary parts of $\gamma$, representing attenuation and dispersion, respectively.

By combining equations for $S_2$ and $S_3$ from (\ref{eqn:s1_s2_s3}) the unknown event signal can be removed from the equations:
\begin{equation}
	\begin{gathered}
		S_3(f) = e^{-(l-2x) \gamma(f)} S_1(f) \\
		S_3(f) = e^{-(l-a) \gamma(f)} S_2(f)
	\end{gathered}
\end{equation}

The measured time signals are transformed into time-frequency domain with the continuous wavelet transform (CWT) \cite{aftabDynamicProtectionPower2020}. This is necessary to locate the TWs in time. The above relation in frequency domain can be written in time-frequency space with CWT \cite{kuleshModelingWaveDispersion2005} as:
\begin{equation}
	\begin{split}
		W&_gs_3(t,f) = \\
		&= e^{-(l-2x)(\alpha(f)+i(\beta(f)-f\beta'(f)))} W_gs_1(t-\frac{(l-2x)\beta'(f)}{2\pi},f) \\
		W&_gs_3(t,f) = \\
		&= e^{-(l-a)(\alpha(f)+i(\beta(f)-f\beta'(f)))} W_gs_2(t-\frac{(l-a)\beta'(f)}{2\pi},f)
	\end{split}
	\label{eqn:wavelet_transform}
\end{equation}
where the transformation from Fourier transform to wavelet transform is done with the approximation of the phase parameter $\beta$ with the first two terms of the Taylor series around $f$. $W$ denotes the continuous wavelet transform and $g$ denotes the mother wavelet. The equation describes that a wave at a specific frequency is delayed, attenuated and phase shifted when passing trough the medium.

We can define a specific point on the measured TWs at the time of the maximum wave amplitude at each wavelet frequency which are defined as $t_i(f)$:
\begin{equation}
	\begin{gathered}
		\max(|W_gs_i(t,f)|)= |W_gs_i(t_i(f),f)| \\
		i \in 1,2,3
	\end{gathered}
	\label{eqn:wavelet_max}
\end{equation}

If we insert the times of maximum wavelet amplitude in (\ref{eqn:wavelet_transform}) in the equation describing the transition of wave trough the line section without the event, we can express the attenuation and dispersion parts of $\gamma(f)$. As the length $l$ is not known with enough accuracy the attenuation and dispersion constants are determined up to the length part. By comparing the delay, real and imaginary parts of (\ref{eqn:wavelet_transform}) we get:
\begin{equation}
	\begin{split}
		&\alpha(f)*l = \frac{ln(\frac{|W_gs_2(t_2(f),f)|}{|W_gs_3(t_3(f),f)|})}{1-\frac{a}{l}} \\
		&\beta'(f)*l = \frac{2\pi\Delta t_{32}(f)}{1-\frac{a}{l}} \\
		&\beta(f)*l = \\
		&= \frac{\angle(W_gs_2(t_2(f),f))-\angle(W_gs_3(t_3(f),f))+2\pi\Delta t_{32}(f)f}{1-\frac{a}{l}}
	\end{split}
\end{equation}
where the $\Delta t_{32}(f)$ is defined as follows by comparing the delay part of (\ref{eqn:wavelet_transform}):
\begin{equation}
	t_3(f) - t_2(f) = \Delta t_{32}(f)=\frac{l-a}{2\pi}\beta'(f)
	\label{eqn:max_t32}
\end{equation}

Similarly $\Delta t_{31}(f)$ can be defined as:
\begin{equation}
	t_3(f) - t_1(f) = \Delta t_{31}(f)=\frac{l-2x}{2\pi}\beta'(f)
	\label{eqn:max_t31}
\end{equation}

By combining the above time differences we can express the event location:
\begin{equation}
	\frac{x}{l}=\frac{1}{2}-\frac{1-\frac{a}{l}}{2}\frac{\Delta t_{31}(f)}{\Delta t_{32}(f)}
	\label{eqn:rel_loc}
\end{equation}

Similarly event location can also be expressed with comparing amplitude or phase shift at the selected time of maximum wavelet amplitude.

The described derivation determines the attenuation and phase parameter of the wave propagation characteristic. The event location with the method in practice is frequency-dependent $\frac{x(f)}{l}$ as the timestamps at each wavelet frequency are influenced by noise so the relative location is processed to remove outliers and averaged. The method was implemented in MatLab and evaluated with simulation which is described in the next section.

\section{Methodology}

This section is divided in two subsections. Subsection~\ref{sec:Simulation} provides the information about the simulation setup. The results of the simulation are generated time-series signals that simulate measurements from measurement devices on a transmission line. The time-series signals were input for the proposed method, which implementation is described in Subsection~\ref{sec:Method}.

\begin{figure}[H]
	\centering
	\includegraphics[clip, width=\figwidth]{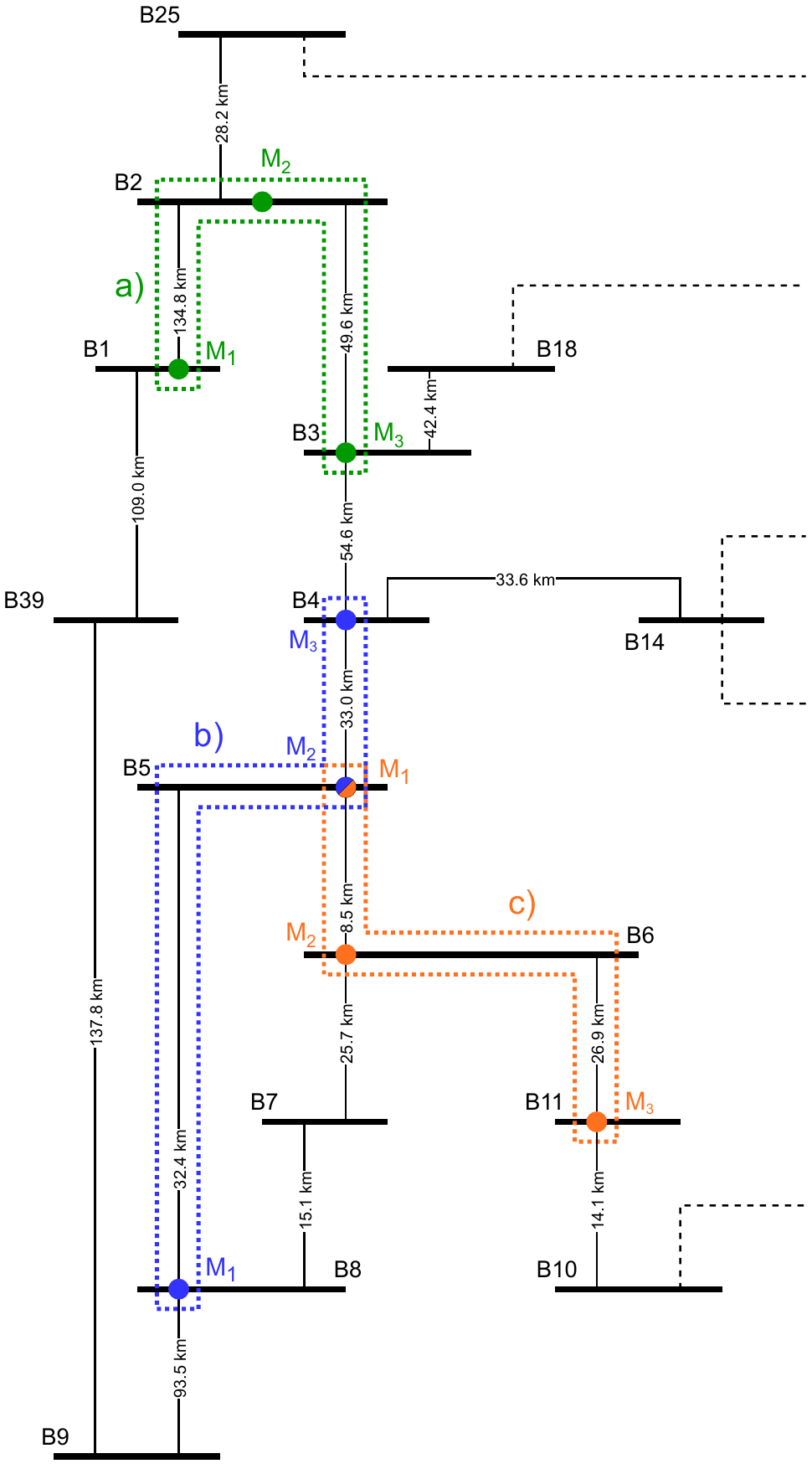}
	\caption{Part of the IEEE 39-bus system with marked line sections of different lengths and locations of measurement devices chosen for evaluation.}
	\label{fig:sim_model}
\end{figure}

\subsection{Transmission line simulation}
\label{sec:Simulation}

The evaluation of the proposed method was performed with the results of electromagnetic transient type simulations. The model is a modified version of the IEEE 39-bus system which specifications are provided in \cite{haddadiPowerSystemTest2018}. Compared to the base IEEE 39-bus model the extended model provides additional information needed for the electromagnetic transient type simulations. It provides the transmission line geometry, conductor types and line lengths and also more details about loads, grid transformers and generators. But the topology of the grid is the same as the base IEEE 39-bus system.

The extended model was implemented in ATPDraw \cite{hoidalenATPDraw} and simulated with ATP-EMTP \cite{ATPEMTP}. Three transmission line sections were chosen from the IEEE 39-bus system that are representative of different line lengths and conditions in the network. The chosen line sections are marked in Fig.~\ref{fig:sim_model} that shows a part of the IEEE 39-bus system. Chosen line sections are between buses B1-B3, B4-B8 and B5-B11.

Line lengths between the buses are of length $l$ as marked in Fig.~\ref{fig:TL_diagram}. The buses connecting two line section parts are B2, B5 and B6 which are of length $a$ from the B1, B8 and B5 buses respectively. On these chosen line sections events were simulated along one part of the line to evaluate the impact of the event location $x$ on the proposed method. The measurement devices were simulated as being located in the buses at each line section ends and the bus in-between which connects both line parts. Both line sections have the same line parameters so the propagation medium does not change along the wave propagation path. The line sections on which the events were generated for evaluation are between buses B1-B2, B5-B8 and B5-B6.

\begin{table}[H]
	\centering
	\small
	\caption{Underground cable model parameters for the frequency-dependent line model.}
	\label{table:line_parameters}
	\SetTblrInner{rowsep=1pt}
	\begin{tblr}{c c}
		\hline
		Parameter & Value \\
		\hline
		Cables depth below ground & 2\,m \\
		Distances between cable phases & 1\,m \\
		Cable core radius & 28.25\,mm \\
		Cable core resistivity & $1.75*10^{-8}\,\Omega$\,m \\
		Internal insulation radius & 47.5\,mm \\
		Metal shield radius & 50.9\,mm \\
		Shield resistivity & $2.8*10^{-7}\,\Omega$\,m \\
		Middle insulation radius & 60.0\,mm \\
		Armour radius & 61.0\,mm \\
		Armour resistivity & $1.68*10^{-8}\,\Omega$\,m \\
		External insulation radius & 70.0\,mm \\
		Relative insulation permittivity & 2.3 \\
		Earth resistivity & 100\,$\Omega$\,m \\
		\hline
	\end{tblr}
\end{table}

To test the method on different transmission line types, the chosen line sections transmission line models were substituted with model representing underground cable. The line parameters of the simulated underground cable are gathered in the Table~\ref{table:line_parameters}. The line lengths stayed the same.

Line sections were modelled with JMarti transmission line model \cite{martiAccurateModellingFrequencyDependent1982}, with 100\,$\Omega$m ground resistance and evaluated in frequency range from 10\,Hz to 10\,MHz with 10 points per decade. The sampling frequency was set to 100\,MHz, which is within the proposed frequency range of 1\,MHz to 1\,GHz proposed by other authors \cite{liangPartialDischargeLocation2021, raoNewCrossCorrelationAlgorithm2020}. The duration of the simulations was set to 1\,ms.

\begin{figure}[H]
	\centering
	\includegraphics[trim=5cm 10.2cm 5cm 10.2cm, clip, width=\figwidth]{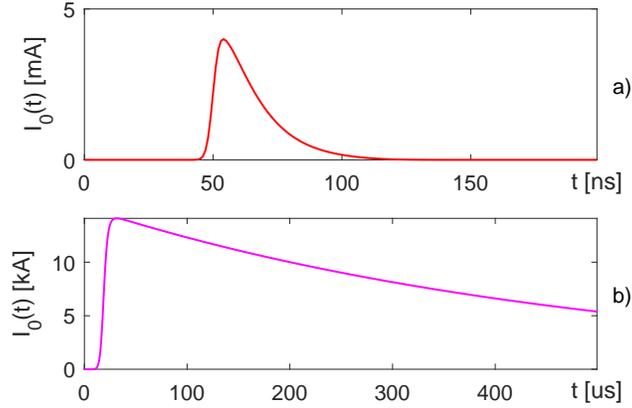}
	\caption{Waveforms of the simulated event types: a) PD and b) lightning strike.}
	\label{fig:sig_waveforms}
\end{figure}

In each simulation experiment the event was generated only on one line section among the ones selected from the IEEE 39-bus model. The events were simulated on a single phase (phase C) with a current generator. The event types simulated were lightning strike and PD. The PD was modelled with double exponent wave \cite{shengNovelOnlineCable2015}, while the lightning strike was modelled with Hiedler's model \cite{furgalInfluenceLightningCurrent2020}. The event waveforms are shown in Fig.~\ref{fig:sig_waveforms}. The lightning strike was simulated in combination with overhead transmission line model and the PD event with the underground cable model.

\begin{table}[H]
	\centering
	\small
	\caption{Evaluated experiments.}
	\label{table:test_cases}
	\SetTblrInner{rowsep=1pt}
	\begin{tblr}{c c c c}
		\hline
		Line type & Event type & Line section & Event location \\
		\hline
		\SetCell[r=3]{m,1.5cm}{Underground cable}& \SetCell[r=3]{m}{PD}& a) & $\frac{n}{10}a; n=1...9$ \\
		& & b) & $\frac{n}{10}a; n=1...9$ \\
		& & c) & $\frac{n}{10}a; n=1...9$ \\
		\hline
		\SetCell[r=3]{m}{Overhead line}& \SetCell[r=3]{m,1.5cm}{Lightning strike}& a) & $\frac{n}{10}a; n=1...9$ \\
		& & b) & $\frac{n}{10}a; n=1...9$ \\
		& & c) & $\frac{n}{10}a; n=1...9$ \\
		\hline
	\end{tblr}
\end{table}

The evaluated experiments are a combination of different event types, line lengths, event locations and transmission line types. The time-series samples from the simulated measurement devices are time synchronised and without noise. The different test experiments parameters are gathered in the Table~\ref{table:test_cases}. To also consider more realistic conditions additional experiments were performed with different noise conditions and when the measurement devices are desynchronized. For the noise evaluation Gaussian noise was added to the raw simulations samples and for the synchronisation experiments a constant time delay was added to the simulation results at two measurement locations.

\subsection{Method implementation}
\label{sec:Method}

\begin{figure}[H]
	\centering
	\includegraphics[width=\figwidth]{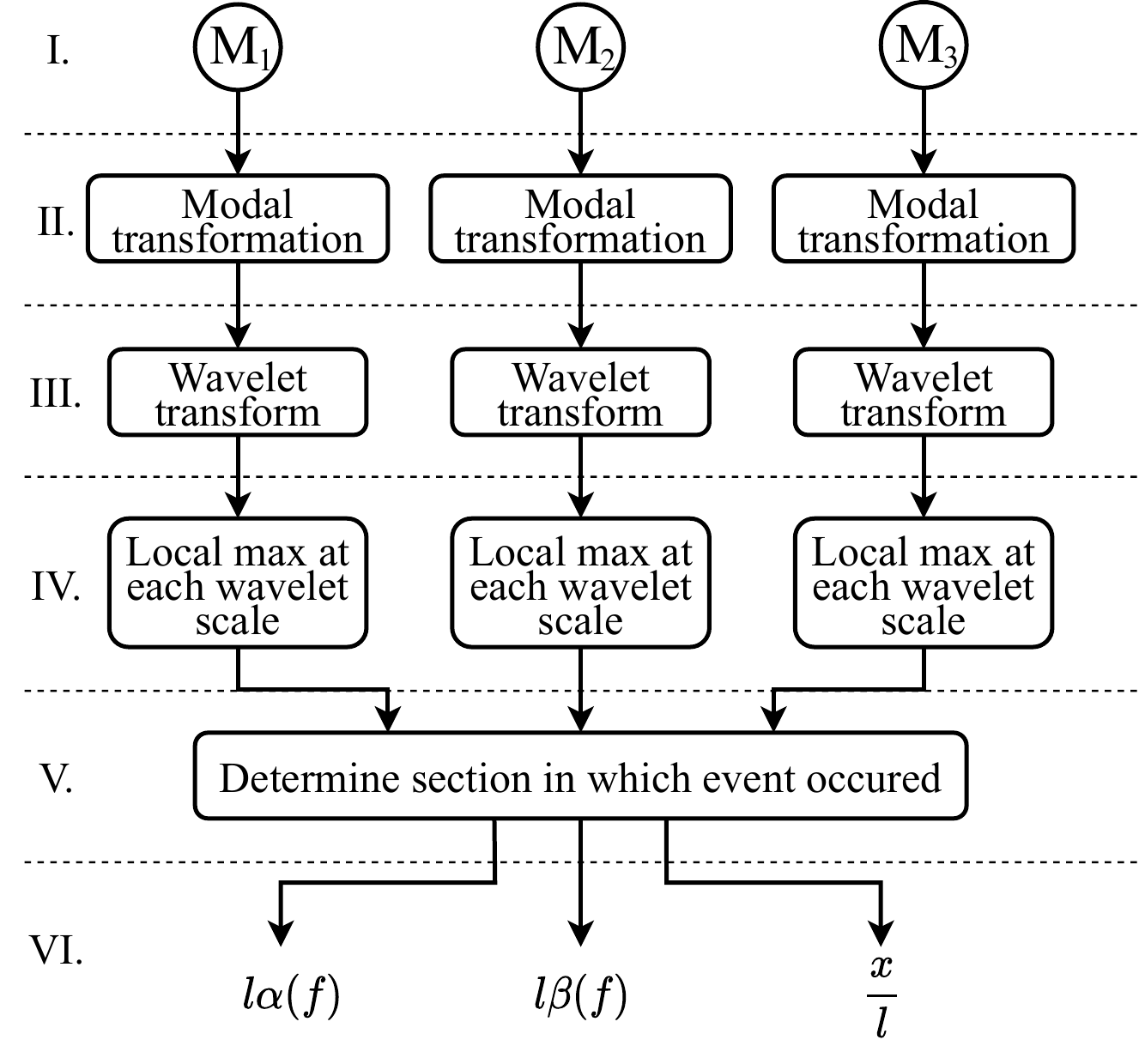}
	\caption{Block diagram of the proposed method implementation.}
	\label{fig:method}
\end{figure}

The proposed method was implemented in MatLab. A block diagram of the method is presented in Fig.~\ref{fig:method}. The diagram shows the different calculation steps, labelled I. through VI., which are described below. For some steps, an instance of the results for one of the experiments are given.

\begin{figure}[H]
	\centering
	\includegraphics[trim=2.2cm 9.3cm 3cm 9.3cm, clip, width=12.75cm]{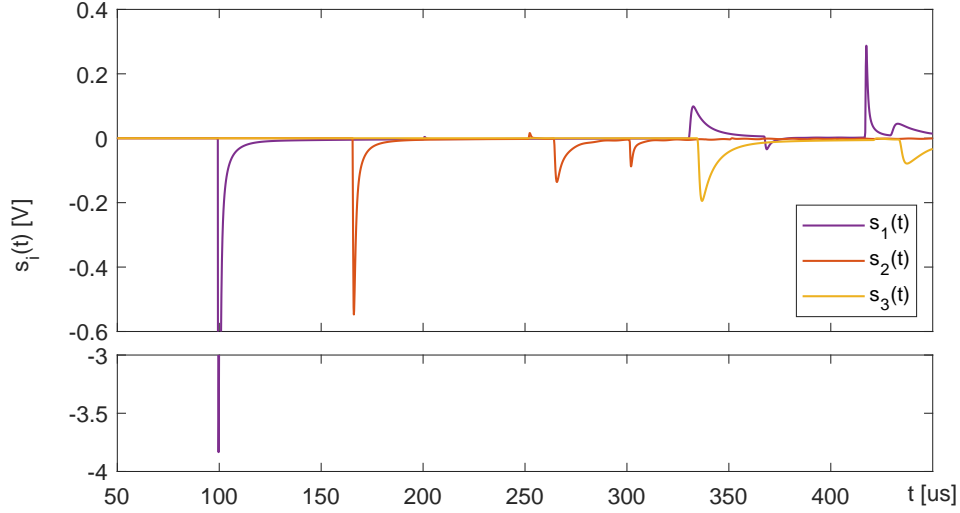}
	\caption{Modal transformed voltage signals for line section b) with underground cable and PD event located at a distance $x=3/10\,a$ after Step II.}
	\label{fig:voltage_signals}
\end{figure}

The time-series samples from the simulation are input to the proposed method which is labelled as Step~I. in Fig.~\ref{fig:method}. An example of the simulation output which was transformed with Clarke matrix to modal values, is shown in Fig.~\ref{fig:voltage_signals}. In the example PD event was generated on the underground cable on the line section b) of the IEEE 39-bus system. The graph in Fig.~\ref{fig:voltage_signals} shows the zero sequence of the transformed measured signals at each measurement location. From the graphs the incident waves as the first spikes can be observed and the reflected waves that arrive after some time. Here the problem of reflected waves in a branched network is shown where it is hard to differentiate from which part of the network the reflection arrived and that the reflected waves are in some cases already quite attenuated.

\begin{figure}[H]
	\centering
	\includegraphics[trim=4.8cm 9.5cm 5.2cm 9.5cm, clip, width=\figwidth]{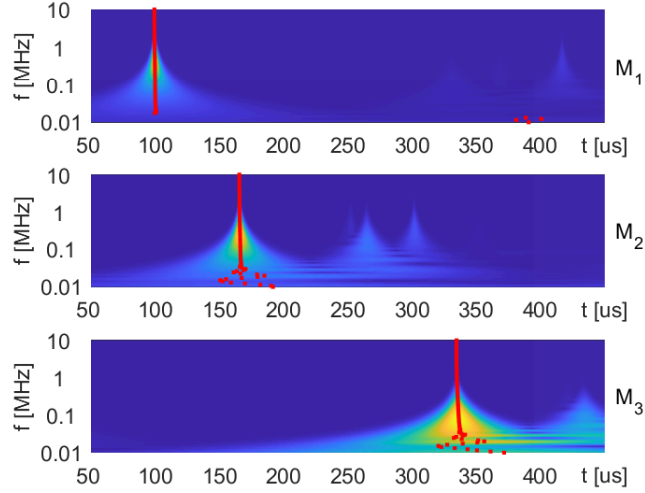}
	\caption{Spectrograms for line section b) with underground cable and PD event located at a distance $x=3/10\,a$ after Step III. and indicated local maximums fo incident waves.}
	\label{fig:cwt_max_graphs}
\end{figure}

The modal samples are transformed with complex Morlet wavelet CWT \cite{kuleshModelingWaveDispersion2005} in Step~III. The wavelet transform gives a complex time-frequency representation of the signal. The time window was 1\,ms as the simulation time. the sampling window In the Step~IV, a local maximum in time is determined in the time-frequency space of the spectrogram representation of the transformed signal at each wavelet scale with the central frequencies $f$. The local maximum acts also as event detection in our case, while in practice another event detection algorithm can be utilized to first identify a TW. An example of Step~III and Step~IV can be found in Fig.~\ref{fig:cwt_max_graphs}, where spectrograms for different measurement locations are shown with indicated maximum values at each wavelet scale.

\begin{figure}[H]
	\centering
	\includegraphics[trim=4.5cm 10cm 5cm 10cm, clip, width=\figwidth]{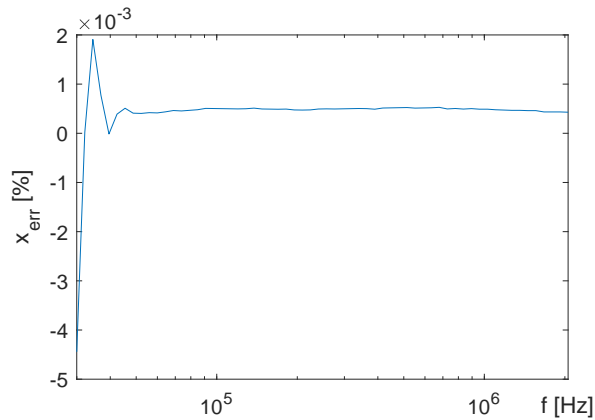}
	\caption{Relative location accuracy for line section b) with underground cable and PD event located at a distance $x=3/10\,a$ after Step VI.}
	\label{fig:n_example}
\end{figure}

In Step~V, the timestamps of the local maximums are compared to determine on which side of the measurement device $M_2$ the event originated. Based on this information the end device of the section where no event was present is selected. By comparing the timestamps of the local maximum at the selected end device and the $M_2$ device, the attenuation parameter, phase parameter and relative location are calculated in Step~VI. An example of relative location error is shown in Fig.~\ref{fig:n_example}. Here, an approximation is made such that the central frequency of each scale represents a single frequency component, as was done in \cite{kuleshModelingWaveDispersion2005}. In this example very accurate result is achieved, where the relative location error is in the range of 0.05\,\% of frequency range between 100\,kHz and 1\,MHz.

After evaluating the implemented method, we performed an accuracy analysis of the method.

\section{Results and discussion}
\label{sec:Results}
The proposed method was evaluated in several use cases based on the simulation setup presented in Section~\ref{sec:Simulation}. First, we compared the method with the state of the art event localization methods which is presented in Section~\ref{sec:Verification}. In Section~\ref{sec:Evaluation} we then evaluated the accuracy of the proposed method in more detail in different experimental setups.

\subsection{Localization comparison}
\label{sec:Verification}
The proposed method was compared with localization results of the existing state of the art methods focusing on methods capable of online localization. We gathered reported localization results for the existing methods and the localization results of our proposed method in Table~\ref{tab:localization_comparisson}. In the table we collected the average and worst reported localization accuracies. The results are presented as relative errors compared to the whole observed line length section with equation:
\begin{equation}
	\begin{gathered}
		x_{error} = \frac{|x_{measured}-x_{model}|}{l_{model}} \\
	\end{gathered}
	\label{eqn:x_relative_error}
\end{equation}

\begin{table}
	\centering
	\fontsize{9}{10}\selectfont
	\caption{Comparison of localization accuracy of different event localization methods.}
	\label{tab:localization_comparisson}
	\SetTblrInner{rowsep=1pt}
	\begin{tblr}{c c c c c c c c}
		\hline
		\SetCell[r=1]{m}Method type & \SetCell[r=1]{m} Reference & Event type & \SetCell[r=1]{m,1.5cm}Lengths evaluated [km] & \SetCell[r=1]{m,1.5cm}Average relative accuracy [\%] & \SetCell[r=1]{m,1.5cm}Worst relative accuracy [\%] & Test setup \\
		\hline
		\SetCell[r=1]{m,2.5cm} classical single-terminal & \cite{liangPartialDischargeLocation2021} & PD & 1.5 & 7.26 & 9.44 & single phase cable \\
		\hline
		\SetCell[r=1]{m,2.5cm} classical double-terminal & \cite{liangPartialDischargeLocation2021} & PD & 1.5 & 1.85 & 4.03 & single phase cable\\
		\hline
		\SetCell[r=2]{m,2.5cm} single-terminal with time reversal & \SetCell[r=2]{c}\cite{ragusaEffectRingMains2022} & \SetCell[r=2]{c} PD & 1.0 & 1.05 & 1.80 & \SetCell[r=2]{c} three phase cable \\
		& & & 2.0 & 0.30 & 0.90 & 0.48 \\
		\hline
		\SetCell[r=1]{m,2.5cm} double single-terminal & \cite{liangPartialDischargeLocation2021} & PD & 1.5 & 0.04 & 0.19 & single phase cable \\
		\hline
		\SetCell[r=1]{m,2.5cm} single-terminal with modelled characteristic & \cite{linTravellingWaveTimefrequency2012} & faults & 400 & 0.15 & 0.45 & HV powered line \\
		\hline
		\SetCell[r=1]{m,2.5cm} multi-terminal with modelled velocity & \cite{chenComputationalFaultTime2019} & faults & $\sim$3 & 0.02 & 0.05 & MV network\\
		\hline
		\SetCell[r=3,c=2]{m,4cm}{Proposed multi-terminal method} & & \SetCell[r=3]{m}{PD, LS} & 184.4 & 0.03 & 0.04 &\SetCell[r=3]{m}{IEEE 39}\\
		& & & 65.4 & 0.01 & 0.03 \\
		& & & 35.4 & 0.05 & 0.08 \\
		\hline
	\end{tblr}
\end{table}

From the compared papers the results were gathered and averaged as reported or recalculated to the same norm of the observed line length. First two methods in Table~\ref{tab:localization_comparisson} are classical single- and double-terminal methods. The results are from \cite{liangPartialDischargeLocation2021} where they compared their proposed method. The methods were tested on shorter cable section which was not a part of a network and had open terminals at the cable ends. The methods were evaluated for PD event. The method proposed in \cite{liangPartialDischargeLocation2021} addresses the changing line propagation parameters by using single ended method from both line ends and iterating the result by varying the propagation velocity so that the locations of the event match from both ends. As such it is not affected by the changing line parameters during operation and it has better localization performance compared to the classical single- and double-terminal methods where the propagation velocity is determined from line model. However, the method has worse performance when event is located near the line terminal where the dispersion effect on the traveling waves in both directions is not the same and with that the method accuracy is affected. Similarly, the method in \cite{chenComputationalFaultTime2019} locates the event based on multiple measurement locations which is extended to a branched network where the measurement locations are located at branch ends. However, only the incident waves are required and the method determines the propagation velocity based on the network section which is not effected by the event. As the method uses the measurements from each branch end the velocity and location result can be refined, but the frequency-dependent wave propagation characteristic is not considered. The method was evaluated on a medium voltage distribution network model. Method presented in \cite{linTravellingWaveTimefrequency2012} is based on continuos wavelet transform modulus maxima, which determines the transient peak in the high frequency spectrum of the signal and is as such more effected by noise and has problem on longer line sections where the attenuation can reduce high frequency components beneath the noise floor level. The method is single-terminal method and uses propagation velocity based on the transmission line model as a setting for the method.

For our method we gathered the results based on the set of simulations reported in Section~\ref{sec:Method}. Results were evaluated in the wavelet central frequency range between 100\,kHz and 1\,MHz. The location was evaluated at each central frequency. From the localization results in the frequency range the outliers were removed and then averaged over the frequency range for the final result. The range was chosen because the signal at higher frequencies in a real environments is already too attenuated, especially on longer lines \cite{mahdipourPartialDischargeLocalization2019}, where the traveling waves recede in the noise. While at lower frequencies, the signal wavelength is longer and the size of the sampling window for the wavelet transform has to be much larger, which is impractical in real-world applications.

Compared to other methods our proposed method determines line propagation characteristic during operation and is as such not effected by medium changes similarly as \cite{chenComputationalFaultTime2019} or \cite{liangPartialDischargeLocation2021}. Besides this the proposed method also considers the dispersion of traveling waves as it evaluates the dispersion at each wavelet central frequency, compared to the other methods that consider the dispersion only indirectly as in the case of \cite{liangPartialDischargeLocation2021} where bigger errors occur with events near line ends or in the case of \cite{linTravellingWaveTimefrequency2012} where the dispersion is evaluated at one frequency component but constant propagation velocity is considered.

To evaluate our method in more detail, we tested it in many different scenarios with a wider range of line lengths and event locations.

\subsection{Method accuracy analysis}
\label{sec:Evaluation}
The proposed method determines the event location and the transmission line propagation characteristic in a frequency range of wavelet central frequencies. Both are dependant on the accuracy of the wavelet transform local maximums timestamps in the frequency range. So in this section, we present the accuracy of the method under different conditions that effect the timestamping.

The conditions effecting the method are different event types, transmission line propagation characteristic, observed line length sections, where the sections are of different lengths and in different parts of the network, event locations along the line section, noise levels and time desynchronization between measurement devices. The simulation experiments done for the analysis are described in Section~\ref{sec:Method}.

The results are presented as the distribution of frequency-dependant localization results. Where relative localization result was determined for each wavelet central frequency by equation:
\begin{equation}
	\begin{gathered}
		x_{error}(f) = \frac{|x_{measured}(f)-x_{model}|}{l_{model}} \\
	\end{gathered}
	\label{eqn:x_relative}
\end{equation}
and then the distribution of the results are presented in graphs where average relative localization errors and standard deviations are marked for each evaluated experiment. The frequency range of the wavelet central frequencies considered is between 100\,kHz to 1\,MHz. As already mentioned, higher frequency components are too attenuated in real conditions and lower frequency components require even longer sampling window for the wavelet transform, which is impractical for the implementation in a real device. The accuracy of localization is proportional to transmission line propagation characteristic accuracy.

\begin{figure}[H]
	\centering
	\includegraphics[trim=0.9cm 10.5cm 0.1cm 10.4cm, clip, width=\linewidth]{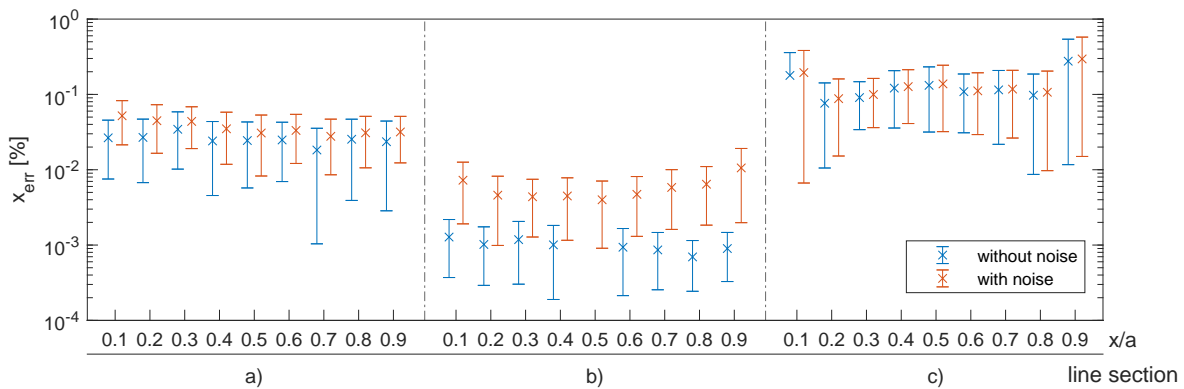}
	\caption{Localization error distribution in the frequency range of 100\,kHz to 1\,MHz for line sections a), b) and c) of overhead transmission lines with lightning strike event.}
	\label{fig:x_res_lightning}
\end{figure}

The first set of experiments were done with lightning strike event Fig.~\ref{fig:sig_waveforms}~b) with the reference overhead transmission lines model, as presented in Section~\ref{sec:Method}. Fig.~\ref{fig:x_res_lightning} shows the distribution of the relative localization errors for different transmission line segments and lighting strike event locations. The results shown include experiments in ideal case without any noise on the transmission line and with added noise. For evaluation under noise conditions, white Gaussian noise was added to the simulation output signals at each measurement location before being processed by the proposed method. The noise was set to a signal-to-noise ratio (SNR) of 60\,dB compared to the peak amplitude value of the event signal. This means that the base noise level was the same for each measurement device and represents the noise floor. With that on longer lines the SNR of the measured signal at the measurement device location was much worse because the TW signal was more attenuated. For the results with noise presented in Fig.~\ref{fig:x_res_lightning} the results were evaluated for 10 simulations with different noise seed.

\begin{figure}[H]
	\centering
	\includegraphics[trim=0.9cm 10.5cm 0.1cm 10.4cm, clip, width=\linewidth]{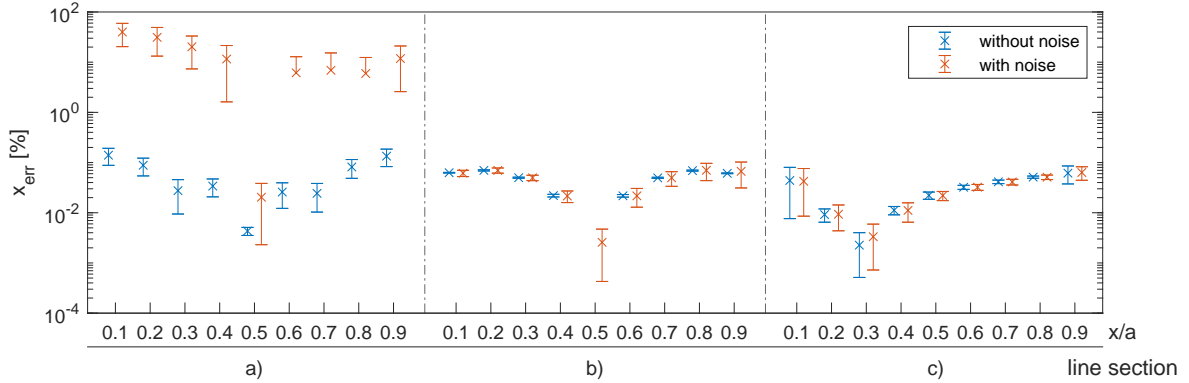}
	\caption{Localization error distribution in the frequency range of 100\,kHz to 1\,MHz for line sections a), b) and c) of underground cable with PD event.}
	\label{fig:x_res_PD}
\end{figure}

The second set of experiments were done with the PD event and transmission line sections replaced with underground cable model presented in Table~\ref{table:line_parameters}. The distribution of the relative localization errors are shown in Fig.~\ref{fig:x_res_PD}. As in the previous case the experiments were done for different transmission line segments, event locations, in ideal and noisy conditions.

The results show that the proposed method works well under different conditions. The relative localization error in the frequency range of 100\,kHz to 1\,MHz is below 1\,\% and mostly even below 0.1\,\%.

In the ideal cases without noise, the measured relative localization error has an error mostly below 0.1\,\%. The variations between the different line sections are because of the different grid topology around the observed line section where the transmissions and reflections in the buses are dependent on the line sections and transformers connected in the buses. The results show that the event localization does not vary much between cases with different event location. The best results are in the middle of the effected line section where the errors cancel out. When event location is near buses, there is some variation in the localization error as seen in experiments with lightning strike event on the line section c). While in experiments with PD event on the line section c) the apparent middle point of the line is skewed because of the transition between the cable and overhead line section. 

Under noise conditions, the relative localization results are within the 1\,\%. The only outliers are the experiments with line section a) with the cable model and PD event in Fig.~\ref{fig:x_res_PD}. In this case the observed transmission line section is over 100\,km long and the TW signal is so attenuated that it falls below the noise floor and the local maximum detection is not successful at the most remote measurement location (bus B3). In other cases the results are a bit worse than the ideal cases.

From the results it is shown that the chosen sampling frequency and frequency range is sufficient but it can be adapted for specific event types, since the slower changing transient events contain less energy at higher frequencies and vice versa. The attenuation and noise limits the high frequency range. The lower frequency limit is set by the wavelet resolution, as it is limited by the sampling time window and the sampling frequency, as well as energy of the transient signals in the lower frequency range.

The measurement devices in the proposed method must be time synchronised if more then one measurement device is deployed. The measurement devices can be synchronized with the Global Positioning System (GPS). The achievable synchronization of the instruments with GPS is better than 100\,ns relative to absolute time \cite{mohamedPartialDischargeLocation2013}. This adds a constant delay to the absolute time error on the measurement devices. Thus, the worst-case time desynchronization between the middle and edge device is $\pm$200\,ns, which causes maximum of 60\,m of location error which amounts to less than 1\,\% in the tested experiments of line lengths. On shorter lines, the timing desynchronization has a greater effect because the time the wave travels between the measurement devices is shorter. For this case of shorter line sections, greater timing accuracy is needed, which can be achieved with optical cables but even with GPS the timing accuracy is typically below 100\,ns. The sampling time jitter that occurs in the event duration is small enough to be considered negligible.

One of the specific parameters affecting our method is the location of the measurement device $M_2$ in a three device setup. From the presented results, it is shown that it does not significantly affect the localization and characterization results. In practice larger distance $b$ between the measurement devices reduces the noise error on the CWT timestamps as the time difference is larger.

An implementation of a measurement device requires a sampling frequency of 10\,MHz to 100\,MHz with support for time synchronization to GPS or synchronization via an optical cable. The input must be compatible with a sensor for voltage or current measurements. Filtering and noise reduction can be performed in the initial signal processing steps to improve signal quality. The real-time processing capabilities for CWT can be implemented in the measurement devices and only the detected local maximum timestamps can be sent to a central location where the rest of the proposed method is implemented. Another option is to deploy other event detection algorithm on the device itself and than transmit a measurement window containing the detected event to a central location where all steps of the proposed method are implemented. This implementation reduces the computational requirements of the measurement devices.

\section{Conclusion}
\label{sec:Conclusion}
In this paper, we propose a new online method for localizing events on transmission lines. The method analyzes the traveling wave in the time-frequency domain of the wavelet transform. Based on three traveling wave measurements, the propagation medium is characterized and the origin of the event is localized. The method thus solves the problem of having to know the wave propagation characteristic of the transmission line in advance. This eliminates the setting as an input parameter for the algorithm, which is one of the limitations of existing methods because the characteristic of the propagation medium changes over time during the operation of the line. At the same time, the characteristic of the transmission line is evaluated in a frequency domain, which improves the localization process by taking into account the dispersion effect.

We have compared the proposed method with other TW methods for event localization, where our method is on par with the best reported methods and has the advantage of accounting for the changing transmission medium over time. We performed a comprehensive analysis of the capabilities of our method in a simulation environment. The method was tested under various conditions in an IEEE 39-bus system model, taking into account different locations in the grid, line lengths, line characteristic, event types, event locations and noise. The results show that the relative localization error is below 0.1\,\% in most cases. The specifics of our proposed method in terms of synchronization requirements were also evaluated. Synchronization has the greatest impact on the accuracy of the method for shorter line sections.

Future work will focus on integrating the proposed method into a test measurement setup and later developing a stand-alone device and the supporting system. In addition, we will investigate where the online transmission line characterization data can be utilized for new services besides improving event localization.

\hfill \break \noindent
\textbf{Funding:} This work was partly funded by the Slovenian Research Agency [grant no. P2-0016], and the European Commission through the H2020 project BD4NRG [grant no. 872613] and Horizon Europe project ENERSHARE [grant no. 101069831]

\bibliographystyle{elsarticle-num}

\end{document}